\documentclass{article}

\usepackage{arxiv}

\usepackage[utf8]{inputenc} 
\usepackage[T1]{fontenc}    
\usepackage{hyperref}       
\usepackage{url}            
\usepackage{booktabs}       
\usepackage{amsfonts}       
\usepackage{nicefrac}       
\usepackage{microtype}      
\usepackage{lipsum}
\usepackage{graphicx}

\usepackage{moreverb}
\usepackage{multirow}
\usepackage{graphicx}
\usepackage{amsmath}
\usepackage{amsthm}
\usepackage{amsfonts}
\usepackage{color}
\usepackage{url}
\usepackage{natbib}

\usepackage{comment}

\newtheorem{theorem}{Theorem}
\newtheorem{lemma}{Lemma}%
\newtheorem{assumption}{Assumption}

\newcommand{\indep}{\perp \!\!\! \perp}
\newcommand{\pr}{\text{Pr}}
\newcommand{\E}{\text{E}}

\title{Causal Effect Estimation on Restricted Mean Survival Time in Case-Cohort Studies via a Matching Design}

\author{
 Andy Ni \\
  Division of Biostatistics\\
  Ohio State University\\
  \texttt{ni.304@osu.edu} \\
   \And
 Wei-En Lu \\
  The Center for Drug Evaluation and Research\\
  Food and Drug Administration\\
  \And
 Bo Lu \\
  Division of Biostatistics\\
  Ohio State University\\
}

\begin{document}
\maketitle
\begin{abstract}
In large observational studies, the case-cohort design is commonly used to reduce the cost associated with covariate measurement. For survival outcomes, literature has suggested that the restricted mean survival time (RMST) be a more appropriate marginal causal effect measure than the hazard ratio. In this paper, we develop a marginal causal effect estimation method for RMST difference under the stratified case-cohort design. We adjust for measured confounders using an innovative template matching design. Compared with conventional matching designs, template matching allows greater flexibility in the sample sizes of the exposed and unexposed groups. We establish the asymptotic properties of the proposed causal effect estimators and develop a bootstrap procedure to estimate their variances. By conducting comprehensive simulation studies, we evaluate the finite sample performance of the proposed estimators, demonstrate the advantage of template matching over conventional matching, and compare between matching on propensity score and matching on covariates. Finally, we apply the proposed methods to the Atherosclerosis Risk in Communities (ARIC) Study to estimate the marginal causal effect of serum hs-CRP level on the coronary heart disease-free survival.
\end{abstract}


\section{Introduction}

\subsection{Causal Inference for Survival Data}
Investigating the causal effects of exposures on the occurrence of clinical events is a central goal in epidemiological and biomedical research. When observational data are used, researchers must address confounding arising from imbalanced covariate distributions between exposed and control groups. Common strategies for confounding adjustment include covariate stratification \citep{rosenbaum1984reducing, cook1989performance, lunceford2004stratification, austin2011introduction, lu2025causal}, covariate matching \citep{rosenbaum1983central, rosenbaum2010design, stuart2010matching, zhao2023flexible}, and inverse probability of treatment weighting \citep{hahn1998role, hirano2001estimation, hirano2003efficient, lunceford2004stratification, imbens2004nonparametric}.

A survival outcome measures the time to an event of interest. Causal inference becomes more challenging when such outcomes are considered. The hazard ratio is a commonly used measure of association in survival analysis. The hazard at a given time is the instantaneous risk of events at that time \textit{among those who are event-free by that time}. The conditional nature of the hazard makes it problematic to draw causal interpretations from a hazard ratio. Specifically, conditioning on survival to time $t$ incurs a selection bias (or collider bias) among survivors at $t$, creating spurious associations between exposure and prognostic covariates, making the latter confounders for the causal effect of exposure on survival risk at time $t+\Delta t$ \citep{hernan2004structural, hernan2010hazards, aalen2015does}. A key manifestation of this selection bias is the noncollapsibility of the hazard ratio \citep{sjolander2016note}, which complicates the estimation of marginal causal effects. 
Robins et al. \cite{robins2000marginal} proposed marginal structural models to estimate marginal causal effects using hazard ratios. However, this approach requires proportional hazards for potential survival times and correct specification of treatment and censoring models over time. These requirements are often difficult to satisfy. Overall, the hazard ratio has well-known limitations as a causal effect measure.

As an alternative to the hazard ratio, the restricted mean survival time (RMST) difference is a useful causal effect estimand. The RMST is the mean survival time during the first $\tau$ units of follow up \citep{irwin1949standard}. It can be computed as the area under the survival curve from time origin to $\tau$. The RMST difference offers several advantages. First, it is free of selection bias, allowing a clear causal interpretation. Second, it is collapsible, facilitating the estimation of both marginal and conditional causal effects. Third, it can be estimated nonparametrically using the Kaplan–Meier estimator and therefore does not rely on model assumptions such as proportional hazards. Finally, the RMST difference directly represents the contrast in expected survival times, making it more intuitive for clinical communication than the hazard ratio. As a result, the RMST difference has gained increasing attention in recent years as a practical and interpretable alternative to the hazard ratio \citep{tian2014predicting, zhao2016restricted, wang2018modeling, zhang2012double, conner2019adjusted, ni2021stratified, lin2023matched}.

\subsection{Case-Cohort Study Design}
Large-scale observational studies often follow thousands of participants for many years to evaluate how exposures affect the occurrence of specific events. In such settings, it may be prohibitively expensive or logistically difficult to measure exposure variables or covariates on every participant, especially when these measurements require costly or invasive bioassays. The case-cohort design was developed to address this challenge in large cohort studies \citep{prentice1986case}. In a case-cohort design, participants who experience the event of interest are termed cases, and those who are right-censored are termed controls. A random subcohort is first selected from the full cohort, and then all cases outside this subcohort are added to it, forming the case-cohort. Expensive exposures or covariates are measured only within this case-cohort, and all subsequent analyses are conducted using the case-cohort sample.

In many observational studies, certain covariates, such as age, sex, and race, are readily available for the entire cohort. To better leverage these variables, Borgan et al. \citep{borgan2000exposure} proposed the \textit{stratified case-cohort design}, in which the full cohort is partitioned into strata based on discretized covariates known for all participants. A case-cohort sample is then drawn separately within each stratum, and the stratum-specific samples are combined to form the overall stratified case-cohort. The sampling probability for the random subcohort may differ across strata. Through simulation studies, Borgan et al. \citep{borgan2000exposure} demonstrated that the stratified case-cohort design can yield more efficient estimation than the standard (non-stratified) case-cohort design.

Because a case-cohort sample is not a representative sample of the full cohort, specialized statistical methods are required for its analysis. In the context of Cox proportional hazards regression, several approaches have been developed to reconstruct the full-cohort risk sets at each event time by applying inverse sampling probability weights to the case-cohort sample \citep{prentice1986case, self1988asymptotic, barlow1994robust, kalbfleisch1988likelihood, borgan2000exposure, kulich2004improving}. Viewing a case-cohort sample as a full cohort with missing exposure or covariate values, Marti et al. \citep{marti2011multiple} proposed an alternative approach based on multiple imputation to analyze case-cohort data.

Research on causal inference under a case–cohort design remains limited. Joffe et al. \citep{joffe1999invited} and Maansson et al. \citep{maansson2007estimation} studied causal inference using propensity scores in case–cohort studies with binary outcomes and odds ratios as the causal effect estimand. Wang et al. \citep{wang2009causal} developed propensity score weighting methods for outcome-dependent two-phase sampling designs with binary and continuous outcomes. Cole et al. \citep{cole2012marginal} and Lee et al. \citep{lee2016marginal} extended marginal structural models to case–cohort designs to estimate marginal causal effects via the hazard ratio, but their approach inherits the limitation of the original marginal structural model of being vulnerable to model misspecification. Most recently, Lu and Ni \citep{lu2025causal} proposed a propensity score stratification method for stratified case–cohort studies with survival outcomes, targeting the marginal causal effect measured by the RMST difference. However, propensity score stratification provides only a coarse level of confounding adjustment. Their simulations showed that substantial residual confounding remains after propensity score stratification.

\subsection{Causal Inference with Matching}
Matching is an important method for confounding adjustment. It pairs subjects from different exposure groups who have similar covariate values, thereby balancing covariate distributions across groups and making the matched comparison resemble a block randomized design \citep{rosenbaum1983central}. Because matching operates at the individual level, it generally achieves substantially better covariate balance than propensity score stratification. Lin et al. \cite{lin2023matched} developed a propensity score matching approach to estimate marginal causal effects using the RMST difference.

When the target estimand is the average treatment effect in the treated (ATT), one approach is to match each treated (\textit{i.e.}, exposed) subject to one or more control (\textit{i.e.}, unexposed) subjects without replacement. This approach requires a substantially larger pool of controls than treated subjects to ensure adequate matching quality—an assumption often violated in real-world datasets. Template matching was developed to address this issue \citep{silber2014template, zhao2023flexible}. In this approach, a representative subset (\textit{i.e.}, template) of treated subjects is first selected, and control subjects are then matched only to this template. This allows the use of a relatively small control group in ATT estimation.

\subsection{Motivating Example: Atherosclerosis Risk In Communities (ARIC) study}
The ARIC study is a prospective cohort study conducted in four U.S. communities to investigate risk factors for the development of cardiovascular diseases \citep{aric1989atherosclerosis}. Because some risk factors, such as high-sensitivity C-reactive protein (hs-CRP), are expensive to measure, the study employed a stratified case-cohort design. The full cohort was stratified by sex, race, and age, and stratum-specific case-cohort samples were drawn and combined to form the stratified case-cohort. Expensive risk factors, including hs-CRP, were measured only within this stratified case-cohort sample. One research question of interest is whether elevated serum hs-CRP causally increases the risk of coronary heart disease (CHD) in the U.S. population with high hs-CRP. In the ARIC case-cohort sample, the high hs-CRP (exposed) group is less than twice the size of the low hs-CRP (unexposed) group, making conventional matching on the treated without replacement likely to yield poor matching quality.

In this paper, we address the above research question by developing template matching methods to estimate marginal causal effects on the treated, using the RMST difference as the estimand, in a stratified case-cohort design with survival outcomes. We focus on one-to-one matching without replacement, employing both propensity score matching and direct covariate matching.

\section{Materials and Methods}

\subsection{Notation and Assumptions}
We consider an exposure variable $A$ with two levels coded as $0$ (unexposed) and $1$ (exposed). Let $T_0^a$ denote the potential survival time under exposure $a=0,1$. Let $X$ be the $p-$dimensional covariate vector and $X_c$ be the $k-$dimensional vector containing only continuous covariates in $X$. Let $H^a(t\mid X)$ be the conditional cumulative hazard function of $T^a$ given $X$. We define the potential \textit{restricted} survival time $T_{0\tau}^a=\min(T_0^a, \tau)$, where $\tau$ is the prespecified truncation time.

The observed survival time $T_0=T_0^1 I(A=1) + T_0^0 I(A=0)$. Similarly, the observed restricted survival time $T_{0\tau}=T_{0\tau}^1 I(A=1) + T_{0\tau}^0 I(A=0)$. Given a random censoring time $C$, the observation time $T=\min(T_{0\tau}, C)$. The conventional definition of events is $\delta=I(T_0<C)I(T_0<\tau)$, where subjects who experience the event of interest before $\tau$ are considered events. For RMST, a more general definition of events is $\delta^*=I(T_{0\tau}<C)$. Under this definition, subjects who survive to $\tau$ without experiencing the event of interest or censoring are also considered as events. The generalized definition of events $\delta^*$ makes intuitive sense because a subject who lives the entire follow-up period $[0, \tau]$ without an event or censoring contributes uncensored full information to the estimation of RMST.

The potential RMST under exposure $a$ is defined as $\mu^a=\E(T_{0\tau}^a)=\int_0^\tau S^a(t)dt$, where $S^a(t)$ is the survival function of the potential survival time $T^a_0$. The potential RMST under exposure $a$ among subjects who receive exposure $A=1$ is defined as $\mu^a_1=\E(T_{0\tau}^a \mid A=1)=\int_0^\tau S^a_1(t)dt$, where $S^a_1$ is the survival function of $T^a_0$ among subjects with $A=1$. The average treatment effect in the treated (ATT) is defined as 
\begin{align}
    \Delta_{ATT}=\mu^1_1(\tau)-\mu^0_1(\tau)=E(T_{0\tau}^1|A=1)-E(T_{0\tau}^0|A=1)=\int^\tau_0\left[S^1_1(t)-S^0_1(t)\right]dt. \notag
\end{align}

Throughout the paper, we make the following two standard assumptions for causal inference \citep{rubin1974estimating}.

\begin{assumption}
\label{sutva}
Stable Unit Treatment Value Assumption (SUTVA). The potential survival times of one individual do not vary with the exposure status of others. And there are no different versions of the specified exposure level. 
\end{assumption}

\begin{assumption}
\label{strong_ignore}
Exposure assignment is strongly ignorable given covariates $X$, that is, $(T_0^0,T_0^1) \indep A \mid X$ and $0<\pr(A=1 \mid X)<1$.
\end{assumption}

Since $\tau$ is a constant, Assumption \ref{strong_ignore} implies that $(T_{0\tau}^0,T_{0\tau}^1) \indep A \mid X$. We make four additional assumptions for the survival data.

\begin{assumption}
\label{indep_cens}
Censoring is independent of potential survival times conditional on exposure status,  $(T_0^0,T_0^1) \indep C \mid A$.
\end{assumption}

This assumption ensures the asymptotic unbiasedness of the marginal survival function estimator in each exposure group in the matched sample. Since $\tau$ is a constant, Assumption \ref{indep_cens} implies $(T_{0\tau}^0,T_{0\tau}^1) \indep C \mid A$.

\begin{assumption}
\label{tau_assump}
The truncation time $\tau$ is smaller than the largest follow up time in both exposure groups.
\end{assumption}

This assumption avoids extrapolation from the largest follow up time to $\tau$, which induces bias in the RMST estimation.

\begin{assumption}
\label{finiteandlipschitzA_assump}
The aboslute value of the continuous covariates $|X_c|<\infty$, $H^a(\tau\mid X)<\infty$ for $a=0,1$ and all $X \in \mathcal{X}$; $H^a(t\mid X_c)$ is Lipschitz with respect to $X_c$ for $a=0,1$ and all $t \in [0, \tau]$.
\end{assumption}

This assumption enforces bounded continuous covariate values and bounded cumulative hazards at $\tau$ for both exposure groups. The Lipschitz condition is required for the bias of the RMST estimator to converge to zero as the matching discrepancy in the covariates converges to zero. This condition is commonly assumed in the literature \citep{abadie2006large, abadie2011bias, savje2022inconsistency}.

\subsection{Stratified Case-Cohort Sampling}

Let $Z$ denote the subset of $X$ that are expensive or difficult to measure. Note that $Z$ can possibly contain the exposure $A$ as well. We consider a stratified case-cohort study with full cohort size $N$ and $B$ strata defined based on covariates that are readily available for the full cohort. Let $\alpha_b$ denote the random subcohort sampling probability in stratum $b$ and let $\xi$ be the subcohort selection indicator. We assume that the subcohorts are selected via Bernoulli sampling to facilitate theoretical development. The resulting stratified case-cohort sample consists of all sampled subcohorts plus all cases outside the subcohorts. Expensive covariates $Z$ (possibly including the exposure $A$) are only observed in the stratified case-cohort sample. Let $n_0$ and $n_1$ denote the unexposed and exposed sample sizes in the stratified case-cohort sample, respectively. Let $n=n_0+n_1$ denote the stratified case-cohort sample size. Additionally, let $N_0$ and $N_1$ denote the unexposed and exposed sample sizes in the full cohort.

For subject $i$, define the case-cohort inverse sampling probability weight as $\rho_i=\delta_i + (1-\delta_i)\xi_i\alpha_{b(i)}^{-1}$, where $b(i)$ indexes the subject's stratum and $\delta=1$ denotes a case. When the causal effect estimand is the RMST difference, one may alternatively select case-cohort samples based on the generalized definition of event $\delta^*=1$. This choice only alters the inverse sampling weights and does not affect the theoretical properties of our proposed methods. We compare the finite sample performance of these two case definitions in the simulation studies. For brevity, we refer to the stratified case-cohort sample simply as the case-cohort sample in the remainder of the paper.

\subsection{Matching on Propensity Scores}

In this section, we develop a propensity score-based template matching method for estimating ATT under a case-cohort design. 

\subsubsection{Template matching on propensity scores}\label{PSmatch}

The first step is to estimate the propensity score $p(X)=\Pr(A=1\mid X)$ by fitting a weighted logistic regression to the case-cohort sample, using the inverse sampling probability weight $\rho$. The stratification variables used in the stratified case-cohort sampling may also be included in the propensity score model. Let $\hat{\beta}$ denote the estimated regression coefficient vector from the weighted logistic regression. The estimated propensity score for the subject $i$ is $\hat{p}(X_i)=\exp(X'_i\hat{\beta})[1+\exp(X'_i\hat{\beta})]^{-1}$. 
We assume the propensity score model is correctly specified. By Proposition 1 of Lu and Ni \cite{lu2025causal}, $\hat{p}(X_i)$ is a consistent estimator of $p(X_i)$ for all $i$.

Next, exposed and unexposed subjects are matched based on their estimated propensity scores. We assume that the unexposed sample size $n_0$ does not substantially exceed the exposed sample size $n_1$ in the case-cohort sample. Under such circumstances, conventional matching methods are unlikely to produce well-balanced matches. To improve matching quality, we use the template matching approach \citep{silber2014template}. Specifically, we first draw a random subset of size $m < n_1$ from exposed subjects to serve as a template representing the exposed group. We then perform one-to-one optimal matching \citep{rosenbaum1989optimal} of unexposed subjects to the template based on propensity scores, resulting in $m$ matched pairs. The reduced number of exposed subjects in the template is expected to improve matching quality. If the template is representative of the original exposure group and the matched pairs achieve good balance, the covariate distributions in both the template and the matched controls resemble those of the exposed population. Hence, we can use the template-matched data to estimate ATT.

To ensure that the template is representative, multiple templates are typically drawn, and the one closest to the original exposed sample is selected for subsequent matching. Various metrics can be used to measure the closeness between the template and the original sample. Under propensity score matching, we propose to measure the closeness by the total pairwise Euclidean distance in propensity scores. Under covariate matching described in Section \ref{Covar_match}, we propose to measure the closeness by the total pairwise Mahalanobis distance. For each candidate template, we compute all $m\times n_T$ pairwise distances between subjects in the template and those in the original exposed group. The template with the smallest total pairwise distances is chosen.

A practical issue in template matching is template size selection. An excessively large template size is likely to result in suboptimal matching quality and, consequently, a biased estimator, whereas an overly small template size can impair the statistical efficiency of the estimator. In the literature, both design-driven \citep{liu2022matching} and data-driven \citep{zhao2023flexible} approaches have been proposed to identify the optimal template size that balances unbiasedness and efficiency. In this paper, we establish the theoretical properties of our matching methods under the assumption that the template size is adequately small to ensure good covariate balance between exposure groups. In the simulation studies, we set the template size to one fifth of the number of unexposed subjects in the case-cohort sample to make sure there are enough subjects in the unexposed pool to match to the template.

\subsubsection{ATT estimation under propensity score matching}\label{ATTPSmatch}

For subject $i$ ($i=1,...,m$), define the counting process $N_i(t)=I(T_i\le t)$ and the at-risk process $Y_i(t)=I(T_i \ge t)$. Suppose that $m$ matched pairs are obtained from template matching on the case-cohort sample. We propose to estimate the cumulative hazard of the potential survival time $T_0^1$ among subjects with $A=1$ using the matched case-cohort sample as
\begin{align}
\hat{H}^1_1(t) &=  \int_0^t\frac{\sum_{i=1}^m dN_{1i}(u)}{\sum_{i=1}^m Y_{1i}(u)\rho_{1i}}, \nonumber
\end{align}
where $N_{1i}(u)$, $Y_{1i}(u)$, and $\rho_{1i}$ are the counting process, the at-risk process, and the case-cohort weight for the exposed subject in the $i$-th matched pair, respectively. $\hat{H}^1_1(t)$ is a weighted version of the nonparametric Nelson-Aalen estimator \citep{nelson1972theory}. 

To estimate the cumulative hazard of the potential survival time $T_0^0$ among subjects with $A=1$, we first estimate the probability of events as a function of the propensity score in the unexposed group of the matched case-cohort sample, $\E_{M_0}[\delta \mid p(X)]$, where $\E_{M_0}(\cdot)$ denotes the expectation with respect to the matched unexposed case-cohort sample. The estimate $\hat{\E}_{M_0}(\delta \mid \hat{p})$ can be obtained from a logistic regression on the matched unexposed case-cohort sample with $\delta$ as the outcome and the estimated propensity score $\hat{p}$ as the sole predictor. Let $\phi(p)=\E_{M_0}[\delta \mid p(X)] + \{1-\E_{M_0}[\delta \mid p(X)]\}\alpha^{-1}$, where $\alpha$ is the random subcohort sampling probability. $\phi(p)$ can be estimated by $\hat{\phi}(\hat{p})=\hat{\E}_{M_0}(\delta \mid \hat{p}) + [1-\hat{\E}_{M_0}(\delta \mid \hat{p})]\alpha^{-1}$. We propose to estimate the cumulative hazard of the potential survival time $T_0^0$ among subjects with $A=1$ using the matched case-cohort sample as
\begin{align}
\hat{H}^0_1(t) &= \int_0^t\frac{\sum_{i=1}^m \rho_{1i}\hat{\phi}(\hat{p}_i)^{-1}dN_{0i}(u)}{\sum_{i=1}^m Y_{0i}(u)\rho_{1i}\rho_{0i}\hat{\phi}(\hat{p}_i)^{-1}}, \nonumber
\end{align}
where $N_{0i}(u)$ and $Y_{0i}(u)$ are the counting process and the at-risk process of the unexposed subject in the $i$-th matched pair, respectively, and $\rho_{0i}$ and $\rho_{1i}$ are the case-cohort weight of the unexposed and exposed subject in the $i$-th matched pair, respectively. 

Following Abadie and Imbens \citep{abadie2006large, abadie2012martingale}, we introduce an additional assumption for the development of the asymptotic properties of the proposed estimators. 
\begin{assumption}
\label{diverge_rate}
$m^r/n_0 \to \theta$ for some $r>k$ and $0<\theta<\infty$.
\end{assumption}

This assumption assumes that, as the sample size tends to infinity, the number of unexposed subjects $n_0$ increases at a faster rate than the number of exposed subjects $m$ in the template. Moreover, the ratio of the divergence rates between unexposed and exposed subjects increases with $k$, the dimension of continuous covariates used in the matching process. This assumption ensures the abundance of unexposed subjects for each exposed subject so that matching discrepancy vanishes as the sample size goes to infinity. 

Let $H^1_1(t)$ and $H^0_1(t)$ be the true cumulative hazards of the potential survival times $T^1_0$ and $T^0_0$ among subjects with $A=1$, respectively. Lemma \ref{lemma1} establishes the asymptotic normality of $\hat{H}^1_1(t)$ and $\hat{H}^0_1(t)$.

\begin{lemma} \label{lemma1}
Under Assumptions \ref{indep_cens} to \ref{diverge_rate}, both $\sqrt{m}[\hat{H}^1_1(t) - H^1_1(t)]$ and $\sqrt{m}[\hat{H}^0_1(t) - H^0_1(t)]$ converge weakly to zero-mean Gaussian processes uniform in $t\in[0,\tau]$ as $m\to\infty$.
\end{lemma}
The proof of Lemma \ref{lemma1} can be found in the Appendix. The estimators of the survival function of the potential survival times $T^1_0$ and $T^0_0$ among subjects with $A=1$ can be derived as $\hat{S}^1_1=\exp[-\hat{H}^1_1(t)]$ and $\hat{S}^0_1=\exp[-\hat{H}^0_1(t)]$, respectively. Thus, the ATT can be estimated by
\begin{align}
    \hat{\Delta}_{ATT}=\int^\tau_0\left[\hat{S}^1_1(t)-\hat{S}^0_1(t)\right]dt. \notag
\end{align}
Theorem \ref{thm1} establishes the asymptotic normality of the causal effect estimator $\hat{\Delta}_{ATT}$.

\begin{theorem} \label{thm1}
Under Assumptions \ref{sutva} to \ref{diverge_rate}, $\sqrt{m}(\hat{\Delta}_{ATT} - \Delta_{ATT})$ converges in distribution to a zero-mean Gaussian variable as $m\to\infty$. 
\end{theorem}
The proof of Theorem \ref{thm1} can be found in the Appendix.

\subsection{Direct Matching on Covariates} \label{Covar_match}

The propensity score matching method relies on the correct specification of the propensity score model. To improve robustness of the matching estimator, one may instead directly match exposed and unexposed subjects on their covariates $X$ using an appropriate distance metric. As with propensity score matching, the stratification variables used in the stratified case-cohort sampling may also be included in this matching process. A widely used distance metric is the Mahalanobis distance \citep{noauthor_reprint_2018}, defined for two realizations $x_1$ and $x_2$ of the matching covariate vector $X$ as
\begin{align}
    d_M(x_1, x_2)=\left[(x_1-x_2)^T \Sigma^{-1} (x_1-x_2)\right]^{1/2}, \notag
\end{align}
where $\Sigma$ is the covariance matrix of $X$. In a case-cohort design, $\Sigma$ can be estimated using case-cohort covariates $X_{cc}$ as
\begin{align}
    \hat{\Sigma}_W=\frac{\sum_{i=1}^n\rho_i(X_{cc,i}-\hat{\mu}_W) (X_{cc,i}-\hat{\mu}_W)^T}{\sum_{i=1}^n \rho_i -1}, \notag
\end{align}
where $\hat{\mu}_W=\sum_{i=1}^n \rho_i X_{cc,i} / \sum_{i=1}^n \rho_i$ is the weighted mean vector of $X_{cc}$, $X_{cc,i}$ is the $i$-th row of $X_{cc}$, $\rho_i$ is the case-cohort weight of subject $i$, and $n$ is the case-cohort sample size.

The same optimal matching algorithm is used to construct the matched sample. The ATT is then estimated using the same procedure as in the propensity score matching method. The only difference is that we estimate the probability of events in the matched unexposed samples $\E_{M_0}[\delta \mid X]$ as a function of covariates instead of the propensity score. $\E_{M_0}[\delta \mid X]$ can be estimated by a logistic regression on the matched unexposed samples with $\delta$ as the outcome and $X$ as the covariates. Subsequently, we calculate $\hat{\phi}(X)=\hat{\E}_{M_0}(\delta \mid X) + [1-\hat{\E}_{M_0}(\delta \mid X)]\alpha^{-1}$. We can establish the same consistency and asymptotic normality of the ATT estimator under covariate matching. The proof is parallel to that of the ATT estimator under propensity score matching, and therefore is not shown.

\subsection{Bootstrap Variance Estimation}

Due to the matching process, the covariates between two matched subjects are more similar than between two unmatched ones. When covariates are associated with survival times, this within-pair similarity of covariates leads to the within-pair similarity of survival times, inducing a correlation of survival times between the two exposure groups in the matched sample. This correlation must be accounted for when estimating the variance of a matching estimator. The case-cohort sampling design further complicates the variance estimation. In this paper, we propose to use a bootstrap approach for variance estimation. 

There is an extensive literature on bootstrap variance estimation for matching estimators. Abadie and Imbens \cite{abadie2008failure} were the first to show that the standard bootstrap that resamples the individual observations is inconsistent for matching \textit{with} replacement. Borody et al. \cite{bodory2024nonparametric} later proposed a martingale-based nonparametric bootstrap method for matching with replacement and established its consistency. For matching \textit{without} replacement, Austin and Small \cite{austin2014use} demonstrated through simulation that the standard bootstrap performs well when the bootstrap resampling is carried out at the level of matched pairs rather than individual observations. Abadie and Spiess \cite{abadie2022robust} formally proved that bootstrap on matched pairs is consistent under matching without replacement. Intuitively, bootstrapping on matched pairs is analogous to an analytical variance estimator that accounts for the clustering of matched pairs. 


Because our study uses matching without replacement, we adopt bootstrap on matched pairs for variance estimation. For a matched case-cohort sample with $m$ pairs, we draw $B$ bootstrap samples of $m$ pairs each, sampled with replacement. Our proposed estimator is applied to each bootstrap sample to obtain $B$ ATT estimates. The original case–cohort weights of the subjects in the bootstrap samples are used to estimate the ATT. The variance of these $B$ bootstrap estimates is taken as the bootstrap variance estimator for the ATT.

\section{Simulation Studies}

\subsection{Data Generation}

We generated six uniformly distributed covariates from a copula with support $[-3, 3]$ and pairwise correlation coefficient of $0.2$. The first three covariates were continuous, and the last three were dichotomized at zero to create binary variables. Thus, the final covariate vector was $X=(X_1, X_2, X_3, X_4, X_5, X_6)$ where $X_1$ to $X_3$ are continuous and $X_4$ to $X_6$ are binary. The exposure variable $A \in {0, 1}$ was generated from a Bernoulli distribution with the parameter $p$ satisfying $\log[p/(1-p)]=\gamma_0 - 0.5 X_1 + 0.5 X_2 - 0.5 X_3 + 0.5 X_4 - 0.5 X_5 + 0.5 X_6$, where $\gamma_0$ controls the marginal proportion of exposure. We considered marginal exposed-to-unexposed ratios of 1:2, 1:3, and 1:4. As the proportion of exposure increases, achieving a well-balanced matched sample becomes more challenging.

The true survival time was generated from an exponential distribution with hazard $h=h_0\exp(1.2 X_1 - 1.2 X_2 + 1.2 X_3 - 1.2 X_4 + 1.2 X_5 - 1.2 X_6 + 3 A + 1.2 X_2 A)$, where $h_0$ is the baseline hazard. An interaction between $X_2$ and $A$ was included so that the marginal causal effect of $A$ differs from its conditional effect. Under this outcome model, the log-hazard ratio of $A$ equals 3 when $X_2=0$ and 4.2 when $X_2=1$. The true ATT in terms of the RMST difference was obtained by generating a very large dataset and empirically calculating the RMST difference between the two potential survival times under $A=0$ and $A=1$.

Censoring times were generated from an exponential distribution independent of $X$ and $A$. The observed survival time was the minimum of the true survival and censoring times. The truncation time was set to the 80-th percentile of the observed survival times. We adjusted the censoring hazard and the baseline hazard $h_0$ to achieve a 10\% event rate under the conventional definition of events and a 30\% event rate under the generalized definition of events. 

We defined four case-cohort strata based on the combination of $X_5$ and $X_6$. From each stratum, a random subcohort was selected via Bernoulli sampling with stratum-specific sampling probabilities determined by the stratum-level event rates. Cases outside the subcohorts were then added to form the stratified case-cohort sample. The resulting case-to-control ratio was close to one in the case-cohort sample. We considered full cohort sample sizes of 5,000 and 10,000.

\subsection{Simulated Data Analysis}

We estimated the ATT using the RMST difference under our proposed template matching method. The template size was chosen so that the ratio of template-to-unexposed was 1:5, which we considered adequate for one-to-one matching on the exposed group. We drew 50 candidate templates from the exposed group and selected the most representative one for subsequent matching based on the total pairwise distance between the original exposed sample and the template, as described in the Materials and Methods section. For comparison, we also estimated the ATT using matching without a template. Both propensity score matching and direct covariate matching using the Mahalanobis distance were implemented. To assess the impact of using the conventional versus generalized definition of events, we generated case-cohort samples under both definitions and compared the performance of the resulting causal effect estimates. 

To estimate the variance of the ATT, we generated 500 bootstrap samples by resampling the matched pairs with replacement and applied the proposed template matching estimator to each bootstrap sample. The variance of the 500 ATT estimates served as the bootstrap estimate of the variance of the ATT.

We conducted 400 replications for each simulation scenario. Percent bias of the estimated causal effect was computed as $100\times (\hat{\Delta}_{ATT}-\Delta_{ATT})/\Delta_{ATT}$. The empirical standard error (SEE) was calculated as the standard deviation of the ATT estimates across simulation replications and was treated as the true standard error. The model-based standard error (SEM) was calculated as the average of the square roots of the bootstrap variance estimates across simulation replications. Empirical coverage probability (CP) was calculated as the proportion of simulation replications in which the 95\% confidence interval for the estimated ATT contained the true ATT. 

\subsection{Simulation Results}
The simulation results for full cohort sample sizes of 5,000 and 10,000 are summarized in Tables \ref{N5000} and \ref{N10000}, respectively. Overall, the bootstrap variance estimator performs quite well, with the model-based standard errors being close to the empirical standard errors across all simulation scenarios. As expected, larger sample sizes reduce variance; however, they do not necessarily lead to a reduction in the bias of the estimated ATT.

\subsubsection{Benefit of using template in matching}
Matching without a template introduces substantial bias (up to $-$20.8\%) in the estimated ATT for both full cohort sample sizes. The bias becomes larger as the unexposed-to-exposed ratio in the original sample decreases, reflecting the impact of deterioration of matching on the accuracy of ATT estimation. The magnitude of the bias is comparable across the two full cohort sizes. Because of the substantial biases in the ATT estimates, their empirical coverages rates fall well below the 95\% nominal level. 

In contrast, template matching greatly reduces bias in the ATT estimates and yields empirical coverage rates close to the nominal level across both full cohort sample sizes and all unexposed-to-exposed ratios in the original sample. As expected, the standard errors of the ATT estimates obtained using template matching are larger than those from matching without a template, since the use of a template reduces the size of the matched sample. This decrease in precision is the trade-off for achieving greater accuracy in the causal effect estimates.

\subsubsection{Impact of definition of events}
When the generalized definition of events is used to construct case-cohort samples, more subjects are classified as cases, resulting in a larger case-cohort sample. Consequently, the ATT estimates exhibit both smaller biases and smaller standard errors compared with those obtained under the conventional definition of events. This pattern holds for both full cohort sample sizes, all unexposed-to-exposed ratios, and both propensity score matching and covariate matching methods.

\subsubsection{Impact of matching methods}
Matching on propensity scores and matching directly on covariates yield comparable levels of bias in the estimated ATT across all scenarios. However, covariate matching consistently produces smaller standard errors than propensity score matching in all scenarios. This may be because covariate matching typically achieves closer covariate values within matched pairs than propensity score matching, inducing stronger within-pair correlation in survival times in the covariate-matched sample than the propensity-score-matched one. As a result, the variance of the estimated ATT is reduced under covariate matching.

\begin{table}[hbt!]
\centering
\caption{Simulation results with N=5,000.  \label{N5000}}%
\begin{tabular}{clrrrrr}
\hline
\textbf{Trt: Control} & \textbf{Method} & \textbf{True Effect} & \textbf{\% bias} & \textbf{SEM} & \textbf{SEE} & \textbf{CP} \\ \hline
& PS matched no template         & & $-$16.7 & 5.8 & 5.9 & 65.8 \\
& PS matched with template          & & 12.7 & 13.1 & 12.9 & 95.8 \\
& Covar matched no template           & & $-$20.8 & 4.9 & 5.4 & 39.5 \\
1:2 & Covar matched with template         & $-$0.0509 & 7.5 & 12.7 & 13.0 & 96.5 \\
& PS matched no template GDE           & & $-$7.7 & 3.5 & 3.3 & 81.3 \\
& PS matched with template GDE        &  & 1.1 & 5.6 & 5.5 & 95.5 \\
& Covar matched no template GDE          &  & $-$3.3 & 3.0 & 3.1 & 91.3 \\
& Covar matched with template GDE        &   & 1.7 & 5.0 & 5.1 & 96.0 \\
\hline
& PS matched no template         & & $-$14.8 & 6.9 & 6.3 & 73.3 \\
& PS matched with template          & & 2.0 & 11.1 & 10.5 & 96.5 \\
& Covar matched no template           & & $-$18.1 & 5.6 & 6.0 & 49.0 \\
1:3 & Covar matched with template         & $-$0.0606 & $-$2.1 & 9.6 & 9.5 & 93.8 \\
& PS matched no template GDE           & & $-$7.7 & 4.5 & 4.1 & 83.5 \\
& PS matched with template GDE       &  & $-$0.59 & 6.0 & 5.8 & 95.3 \\
& Covar matched no template GDE             &  & $-$2.9 & 3.8 & 3.8 & 92.8 \\
& Covar matched with template GDE       &  & 0.96 & 5.1 & 5.2 & 95.5 \\
\hline
& PS matched no template         & & $-$10.5 & 7.6 & 7.2 & 86.6 \\
& PS matched with template          & & 6.7 & 12.3 & 11.1 & 97.0 \\
& Covar matched no template           & & $-$16.6 & 6.6 & 7.2 & 66.5 \\
1:4 & Covar matched with template         & $-$0.0542 & $-$2.5 & 10.8 & 10.8 & 93.0 \\
& PS matched no template GDE            & & $-$4.5 & 4.5 & 4.0 & 94.3 \\
& PS matched with template GDE       &  & $-$0.79 & 5.4 & 4.8 & 97.0 \\
& Covar matched no template GDE            &   & $-$1.8 & 4.0 & 3.9 & 96.3 \\
& Covar matched with template GDE       &  & $-$0.18 & 4.7 & 4.8 & 94.0 \\
\hline
\end{tabular}\\

\begin{footnotesize}
    PS: propensity score; Covar: covariate; GDE: generalized definition of events; SEM: model-based standard error; SEE: empirical standard error; CP: 95\% empirical coverage.
\end{footnotesize}
\end{table}

\begin{table}[hbt!]
\centering
\caption{Simulation results with N=10,000.  \label{N10000}}%
\begin{tabular}{clrrrrr}
\hline
\textbf{Trt: Control} & \textbf{Method} & \textbf{True Effect} & \textbf{\% bias} & \textbf{SEM} & \textbf{SEE} & \textbf{CP} \\ \hline
& PS matched no template         & & $-$17.4 & 4.1 & 3.9 & 40.5 \\
& PS matched with template        & & 6.7 & 8.6 & 8.7 & 95.5   \\
& Covar matched no template          & & $-$19.6 & 3.3 & 3.8 & 17.3  \\
1:2 & Covar matched with template         & $-$0.0509 & 6.4 & 7.9 & 7.5 & 96.3 \\
& PS matched no template GDE           & & $-$7.8 & 2.5 & 2.4 & 64.3 \\
& PS matched with template GDE        & & 0.68 & 4.0 & 4.1 & 95.0 \\
& Covar matched no template GDE          & & $-$3.0 & 2.1 & 2.4 & 85.7 \\
& Covar matched with template GDE        & & 1.7 & 3.5 & 3.4 & 94.8 \\
\hline
& PS matched no template         & & $-$14.8 & 6.9 & 6.3 & 73.3 \\
& PS matched with template          & & 2.0 & 11.1 & 10.5 & 96.5 \\
& Covar matched no template           & & $-$18.1 & 5.6 & 6.0 & 49.0 \\
1:3 & Covar matched with template         & $-$0.0606 & $-$2.1 & 9.6 & 9.5 & 93.8 \\
& PS matched no template GDE           & & $-$7.2 & 3.2 & 3.0 & 72.3 \\
& PS matched with template GDE       & & $-$1.2 & 4.2 & 3.9 & 96.5 \\
& Covar matched no template GDE             & & $-$2.1 & 2.7 & 2.7 & 92.7 \\
& Covar matched with template GDE       & & 1.5 & 3.6 & 3.6 & 93.4 \\
\hline
& PS matched no template         & & $-$10.5 & 5.3 & 5.4 & 80.5 \\
& PS matched with template          & & 3.0 & 8.0 & 8.1 & 95.3 \\
& Covar matched no template           & & $-$14.4 & 4.4 & 4.9 & 52.8 \\
1:4 & Covar matched with template         & $-$0.0542 & $-$0.57 & 7.1 & 7.1 & 94.3 \\
& PS matched no template GDE            & & $-$5.0 & 3.2 & 2.9 & 88.3 \\
& PS matched with template GDE       & & $-$2.5 & 3.8 & 3.6 & 94.3 \\
& Covar matched no template GDE            & & $-$1.9 & 2.8 & 2.6 & 94.8 \\
& Covar matched with template GDE       &  & $-$0.59 & 3.3 & 3.0 & 96.3 \\
\hline
\end{tabular}\\

\begin{footnotesize}%
PS: propensity score; Covar: covariate; GDE: generalized definition of events; SEM: model-based standard error; SEE: empirical standard error; CP: 95\% empirical coverage.
\end{footnotesize}
\end{table}

\section{Real Data Analysis}

We applied the proposed matching methods to the ARIC study to estimate the ATT of hs-CRP on the RMST of CHD-free survival. Enrollment for the study began between 1987 and 1989, and 15,792 participants were followed through 1998 \citep{ballantyne2004lipoprotein}. Individuals who remained CHD-free by the end of 1998 or were lost to follow-up were treated as censored. Among the full cohort, 638 developed CHD or died, yielding a censoring rate of 94.8\%. For this analysis, we used the 12,197 participants with complete baseline covariate data. 

The study investigators selected a stratified case-cohort sample of 1,567 participants, using strata defined by sex, race (black vs. white), and baseline age ($\le 55$ vs. $>55$). To reduce costs, the investigators only measured hs-CRP within the case-cohort sample. Since the case-cohort sample was selected using the conventional definition of events, it is not possible to redraw a new case-cohort sample based on the generalized definition of events and obtain hs-CRP measurements for it. 

We dichotomized the hs-CRP into high ($>3$ mg/L) and low ($\le 3$ mg/L) levels, treating high hs-CRP level as the exposure. Thus, the ATT represents the average causal effect of elevated hs-CRP among participants with high hs-CRP. Given the follow-up duration, we selected a truncation time $\tau$ of 8 years (2,920 days) for computing the RMST. 

We included ten common prognosticators of CHD in the analysis. Their summary statistics in the case–cohort sample by hs-CRP level are presented in Table \ref{Real data summary stat}. These statistics reveal substantial covariate imbalances between the two hs-CRP groups, suggesting that many variables may confound the causal effect of hs-CRP on the RMST of CHD-free survival. The unexposed-to-exposed ratio in this study is $947/620\approx 1.5$, which is too low to support high-quality matching, making template matching necessary. 

We set the template size to 190 to achieve the same unexposed-to-exposed ratio of five as in the simulation studies. As a sensitivity analysis, we also used a template size of 240, representing an unexposed-to-exposed ratio of four. From the exposed group, we drew 75 random candidate templates and selected the one with the smallest total pairwise distance to the original exposed sample. Distances were computed using either Euclidean distances in propensity scores or Mahalanobis distances in covariates, depending on the matching method. Optimal matching was then carried out using the selected template.

\begin{figure}
    \centering
    \includegraphics[width=13cm]{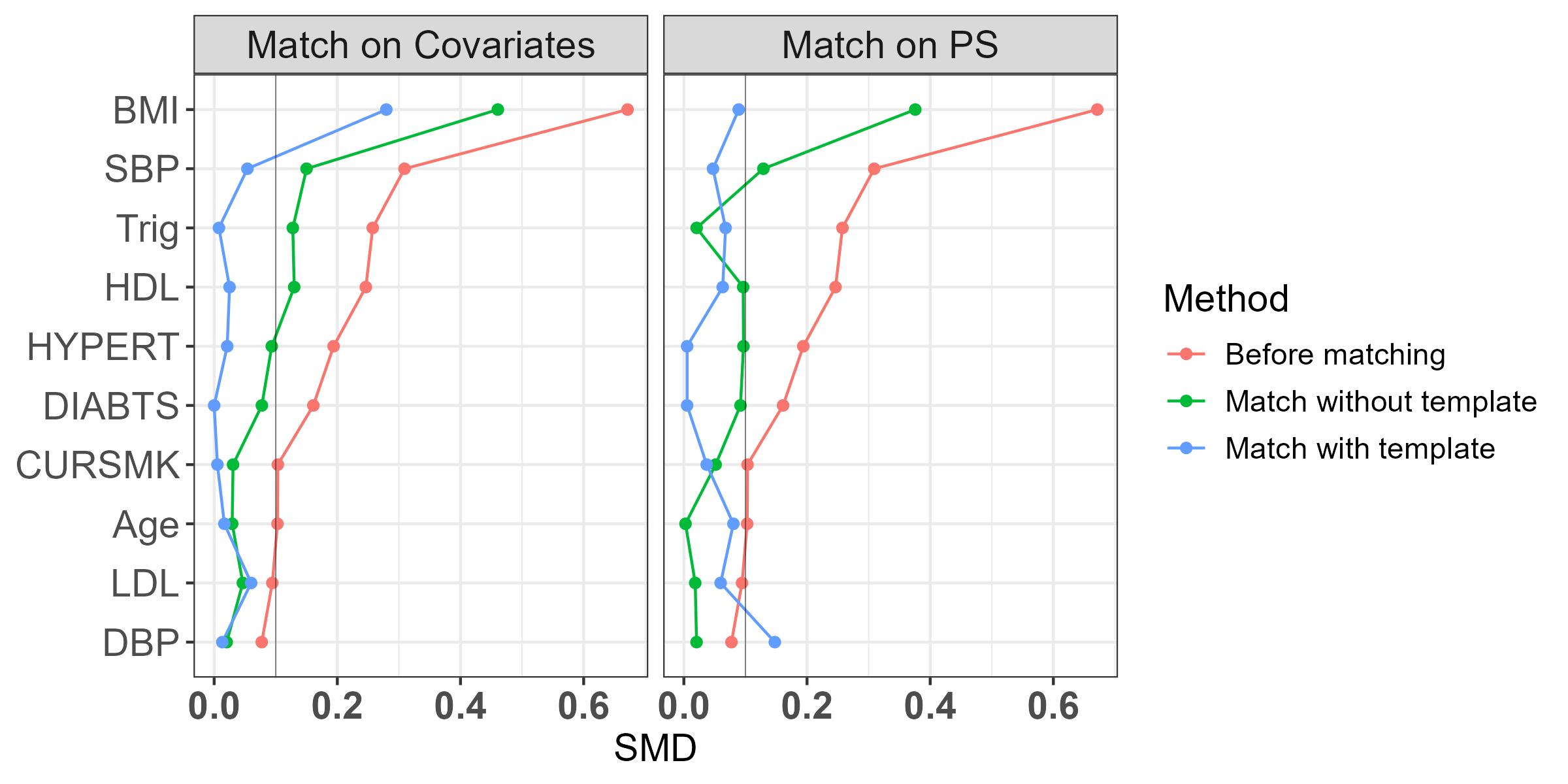} 
    \caption{Standardized mean differences of covariates used in the ARIC study analysis before and after matching.}
    \label{BalancePlot}
\end{figure}

We evaluated covariate balance between the two hs-CRP groups before and after matching, both with and without templates. For continuous covariates, balance was quantified using the standardized mean difference (SMD), defined as $\text{SMD}=(\mu_1 - \mu_0)/\sigma_{pool}$, where $\sigma_{pool}$ is the pooled standard deviation in the case-cohort sample. For categorical covariates, balance was assessed using the differences in proportions. Figure \ref{BalancePlot} summarizes the covariate balance under no matching, matching without a template, and matching with a template of size 190. Matching, whether on covariates or propensity scores, substantially improved covariate balance between the two hs-CRP groups. For covariate matching, template matching consistently produced better balance than matching without a template. For propensity score matching, template matching generally improves balance, though not uniformly across all covariates. With a template, covariate matching achieved better balance than propensity score matching for all covariates except BMI.

We then applied the proposed ATT estimation methods to the matched samples. Table \ref{ARICresults} summarizes the estimated ATTs (in days) of hs-CRP on the RMST of CHD-free survival, along with their standard errors and 95\% confidence intervals, for both propensity score matching and covariate matching, with and without templates. Without a template, ATT estimates differ markedly between propensity score matching and covariate matching (2.3 vs. $-$26.5 days), indicating that ATT estimates can be unstable when matching is performed without a template. In contrast, when a template is used to increase the unexposed-to-exposed ratio to five, the two matching methods yield much more similar estimates ($-$18.2 vs. $-$13.2 days). As expected from the bias–variance tradeoff, standard errors are larger for template-matched estimates than for those from matching without a template. Consistent with the simulations, covariate matching produces smaller standard errors than propensity score matching.

The template-matched ATT estimates from covariate and propensity score matching methods are both negative, suggesting a potentially harmful effect of elevated hs-CRP on the RMST of CHD-free survival. However, the 95\% confidence intervals for both estimates include zero. Therefore, at the 5\% significance level, there is insufficient evidence to conclude that elevated serum hs-CRP causally affects CHD-free RMST among individuals with high hs-CRP. In the sensitivity analysis, using an unexposed-to-exposed ratio of four yields ATT estimates that are reasonably close to those reported in Table \ref{ARICresults}, with the same conclusions. This suggests that the results are fairly robust to the template size in this dataset.

\begin{table}[hbt!]
    \centering
    \caption{\label{Real data summary stat}Summary statistics of baseline variables by hs-CRP protein level between 0 and 3 mg/L and greater than 3 mg/L.}
    \begin{tabular}{lll}
    \hline
    \textbf{Variable}               & \textbf{\begin{tabular}[c]{@{}l@{}}hs-CRP $\leq$ 3 mg/L \\ Mean (SD) or n (\%) \end{tabular}} & \textbf{\begin{tabular}[c]{@{}l@{}}hs-CRP \textgreater 3 mg/L\\ Mean (SD) or n (\%) \end{tabular}} \\ \hline
    Sample size & 947 & 620 \\ \hline
    Age & 58.16 (5.59) & 58.73 (5.42) \\
    Number of current smoker & 203 (21.4) & 197 (31.8) \\
    Number of diabetic participants & 156 (16.5) & 202 (32.6) \\
    BMI & 27.00 (4.20) & 30.42 (6.23) \\
    LDL (mg/dL) & 138.05 (37.82) & 141.67 (39.06) \\
    HDL (mg/dL) & 48.41 (16.00) & 44.59 (14.73) \\
    Triglyceride (mmol/L) & 128.40 (65.06) & 145.45 (67.88) \\    
    Systolic blood pressure (mmHg) & 124.22 (19.25) & 130.40 (21.06) \\
    Diastolic blood pressure (mmHg) & 73.46 (11.07) & 74.30 (10.87)  \\  
    Number of Hypertensive participants & 354 (37.4) & 352 (56.8) \\
    \hline
    \end{tabular}
\end{table}

\begin{table}[hbt!]
\centering
\caption{Estimated ATT (in days) of hs-CRP on RMST of CHD-free survival in ARIC study using matching without a template and with templates to achieve unexposed-to-exposed ratios of five and four.  \label{ARICresults}}%
\begin{tabular}{lccc}
\hline
\textbf{Method} & \textbf{ATT} & \textbf{SE} & \textbf{95\% CI} \\ \hline
PS matched no template         & 2.3 & 17.5 & $-$26.6 to 37.8 \\
PS matched with template & $-$18.2 & 25.7 & $-$71.2 to 32.5   \\
Covar matched no template          & $-$26.5 & 12.5 & $-$48.7 to $-$1.94  \\
Covar matched with template         & $-$13.2 & 24.3 & $-$63.0 to 31.8 \\
\hline
\end{tabular}
\end{table}

\section{Discussion}

In this paper, we develop matching methods to estimate the average treatment effect in the treated (ATT) on the restricted mean survival time (RMST) under a stratified case–cohort study design. We consider both propensity score matching and direct covariate matching. The theoretical development of these estimators is complicated by the case–cohort sampling scheme. Using empirical process techniques, we establish the consistency and asymptotic normality of the proposed estimators under suitable assumptions. We use the bootstrap to estimate the variances of the matching estimators. Simulation studies further illustrate the finite-sample performance of the proposed methods.

Our simulation study shows that matching on covariates yields more precise estimates than matching on propensity scores. The ARIC data analysis further demonstrates that covariate matching achieves better covariate balance than propensity score matching. Together, these findings suggest that matching directly on covariates provides higher quality matches, and therefore more reliable causal effect estimates than matching on propensity scores.

However, a key limitation of covariate matching is that it becomes inefficient or even infeasible in high-dimensional settings due to the curse of dimensionality and increased computational burden. In contrast, propensity score matching remains practical in high-dimensional scenarios by reducing all covariates to a single scalar, and it generally provides reasonably good causal effect estimates. For this reason, we recommend using covariate matching whenever the dimensionality of the covariates permits, and relying on propensity score matching for high-dimensional data.

A third option is a hybrid approach that combines the strengths of both methods. When a key subset of covariates is of particular scientific interest or especially difficult to balance, one can match directly on this subset while imposing propensity score calipers, where the propensity score is estimated from all covariates \citep{rubin2000combining}. In this approach, the distance between two subjects is defined as their Mahalanobis distance if the absolute difference in their propensity scores is no greater than a caliper $c$; otherwise, the distance is set to infinity. This strategy typically achieves excellent balance for the key covariates and reasonably good balance for the remaining covariates. It is important to note that such a calipered distance should not be used when evaluating the representativeness of a template because it assigns an infinite distance to many pairs, leading to infinite total pairwise distances. Instead, Mahalanobis distance or the Euclidean distance of the propensity scores should be used for template evaluation, even if calipered distances are used to construct the matched samples.

Assumption \ref{diverge_rate} requires that the numbers of exposed and unexposed subjects diverge to infinity at different rates. This implies that the proportion of exposed subjects in the population diminishes as the size of the population goes to infinity. This setting differs from the conventional asymptotic framework, in which the population composition is assumed to remain fixed along the asymptotic sequence. Nevertheless, as discussed by Savje \citep{savje2022inconsistency}, such an assumption is reasonable in many practical scenarios. For example, if the exposure of interest is a novel treatment, it is natural to expect substantially more unexposed subjects than exposed ones in the population. Moreover, the template matching procedure proposed in this paper further ensures that the number of unexposed subjects is sufficiently large relative to the number of exposed subjects in the template, thereby enhancing the plausibility of Assumption \ref{diverge_rate}.

Although this paper focuses on estimating the ATT, template matching can also be employed to estimate the overall average treatment effect (ATE). Recently, Zhao and Lu \citep{zhao2023flexible} proposed a triplet matching algorithm that matches exposed and unexposed subjects to a common template that represents the entire population, enabling ATE estimation using the resulting matched sample. The estimators proposed in this paper can be extended to this setting. Under triplet matching, not all exposed subjects are necessarily included in the matched sample. Consequently, the cumulative hazard estimator for the exposed group, $\hat{H}^1_1(t)$, would take a form analogous to that of the unexposed group estimator, $\hat{H}^0_1(t)$. 

Another natural extension of the proposed method is to consider matching \textit{with} replacement, where an unexposed subject may be matched to multiple exposed subjects. Matching with replacement eliminates the need for a template. However, variance estimation becomes substantially more challenging in this setting, as the standard bootstrap is inconsistent under matching with replacement \citep{abadie2008failure, abadie2022robust}. To obtain valid variance estimates, one would need to either derive an analytical variance formula that accounts for the case-cohort sampling scheme and the correlation among matched pairs induced by repeatedly using the same unexposed subjects, or to adapt the martingale-based bootstrap method of Borody et al. \cite{bodory2024nonparametric} to the case-cohort design.

\section{Acknowledgments}
This work was partially supported by grant 1R21HL170212 from National Heart, Lung, and Blood Institute. The authors thank the staff and participants of the ARIC study for their important
contributions. The ARIC Study is carried out as a collaborative study supported by National Heart, Lung, and Blood Institute contracts (N01-HC-55015, N01-HC-55016, N01-HC-55018, N01-HC-55019, N01-HC-55020, N01-HC-55021, N01-HC-55022).

\bibliographystyle{unsrt}  
\bibliography{reference}

\appendix

\section{Proof of Lemma 1}

We first establish the asymptotic distribution of $\sqrt{m}[\hat{H}^1_1(t) - H^1_1(t)]$. 

\begin{align*}
    & \sqrt{m}[\hat{H}^1_1(t) - H^1_1(t)] \\
    & = \sqrt{m}\left[\int_0^t \frac{\sum_{i=1}^m dN_{1i}(u)}{\sum_{i=1}^m Y_{1i}(u) \rho_{1i} }- H^1_1(t) \right]\\
    & = \int_0^t \frac{m}{\sum_{i=1}^m Y_{1i}(u) \rho_{1i}} \frac{1}{\sqrt{m}} \sum_{i=1}^m dM_{1i}(u) + \sqrt{m} \int_0^t \frac{\sum_{i=1}^m Y_{1i}(u) - \sum_{i=1}^m Y_{1i}(u) \rho_{1i}}{\sum_{i=1}^m Y_{1i}(u) \rho_{1i}} dH^1_1(u),
\end{align*}
where $dM_{1i}(u) = dN_{1i}(u) - Y_{1i}(u)dH^1_1(u)$ is a martingale by Doob Meyer decomposition.

Let $\mathcal{F}_1(t)$ be the sigma algebra generated by  $N_{1i}(u)$, $Y_{1i}(u)$, and $\delta_{1i}$ for $0 \leq u \leq t$ and $i=1,...,m$. Then $\E\big[\xi_{1i} \mid \mathcal{F}_1(t)\big]=\alpha_{b(i)}$ where $\xi_{1i}$'s are the random subcohort selection indicator. Under Bernoulli sampling, they are i.i.d. By the weak law of large numbers, $m^{-1}\sum_{i=1}^m Y_{1i}(u)\rho_{1i} \xrightarrow{p} \pi_1(u)$ where $\pi_1(u)=\E[Y_{1i}(u)\rho_{1i}]= \E[Y_{1i}(u)\E(\rho_{1i} \mid \mathcal{F}_1(t))] = \E[Y_{1i}(u)]$ since $\E[\rho_{1i} \mid \mathcal{F}_1(t)] = 1$. It follows that
\begin{align*}
    & \sqrt{m} [\hat{H}^1_1(t) - H^1_1(t)] \\
    & = \int_0^t \pi_1(u)^{-1}\frac{1}{\sqrt{m}}\sum_{i=1}^m dM_{1i}(u) + \frac{1}{\sqrt{m}} \int_0^t \frac{\sum_{i=1}^m Y_{1i}(u)(1-\rho_{1i})}{\pi_1(u)} dH^1_1(u) + o_p(1) \\
    & = I_1 + I_2 + o_p(1).
\end{align*}

Let us first consider the term $I_1$. Since $M_{1i}(u)$ is a martingale and $\pi_1(u)^{-1}$ is a bounded deterministic process, by the Martingale Central Limit Theorem, $I_1$ converges weakly to a mean-zero Gaussian process on $[0,\tau]$.

Now let us consider the term $I_2$. Under Bernoulli sampling, $Y_{1i}(u)(1-\rho_{1i})\pi_1(u)^{-1} dH^1_1(u)$ are i.i.d. random processes with bounded variation. Applying the Central Limit Theorem at each time $t$, we know that $I_2$ converges pointwise in distribution to a normal distribution with mean $\E(I_2)=\E[\int_0^t Y_{1i}(u)(1-\rho_{1i})\pi_1(u)^{-1}dH^1_1(u)]=0$ since $\E[Y_{1i}(u)\rho_{1i}]=\E[Y_{1i}(u)]=\pi_1(u)$. By Example 2.5.4 in Vaart and Wellner \citep{wellner2013weak}, the class of indicator functions $\{Y_{1i}(u), u\in[0,t]\}$ is Donsker. Since $1-\rho_{1i}$, $\pi_1(u)^{-1}$, and $dH^1_1(u)$ are all bounded, it follows that $Y_{1i}(u)(1-\rho_{1i})\pi_1(u)^{-1} dH^1_1(u)$ is Donsker and therefore tight on $[0,t]$. Thus, $I_2$ converges weakly to a mean-zero Gaussian process.

Furthermore, since $\E(I_1)=0$ and $\E(I_2)=0$,
\begin{align*}
    \mathrm{cov}(I_1,I_2) & = \mathrm{E}(I_1 I_2) \\
    & = \mathrm{E}\big[I_1 \mathrm{E}(I_2\mid \mathcal{F}_1(t)) \big] \\
    & = \mathrm{E}\Bigg[I_1 \int_0^t \frac{1}{\sqrt{m}} \sum_{j=1}^m Y_{1i}(u)(1-\delta_{1i})\left(1-\frac{\E(\xi_{1i} \mid \mathcal{F}_1(t))}{\alpha_{b(i)}}\right)\pi_1^{-1}(u) dH^1_1(u) \Bigg] \\
    & = 0.
\end{align*}
Therefore, $I_1$ and $I_2$ are independent of each other. Putting the above results together, $\sqrt{m} [\hat{H}^1_1(t) - H^1_1(t)]$ converges weakly to a mean-zero Gaussian process on $[0, \tau]$.

Next, we establish the asymptotic distribution of $\sqrt{m}[\hat{H}^0_1(t) - H^0_1(t)]$. Since $\hat{H}^0_1(t)$ is based on the matched unexposed subjects but $H^0_1(t)$ is for the exposed subjects had they not been exposed, any discrepancy in the propensity scores between matched subjects will lead to bias in $\hat{H}^0_1(t)$ \citep{abadie2006large, abadie2011bias, savje2022inconsistency}. We first show that this bias due to matching discrepancy goes to zero under the assumption that $m^r/n_0 \to \theta$ for some $r>k$ and $0<\theta<\infty$. The empirical marginal cumulative hazard based on the matched unexposed subjects can be written as $\sum_{i=1}^m H^0(t \mid p_{0i})S^0(t \mid p_{0i})/\sum_{i=1}^m S^0(t \mid p_{0i})$. Under the assumption that $m^r/n_0 \to \theta$ for some $r>k$ and $0<\theta<\infty$, $n_0$ goes to infinity faster than $m$, so there will be infinitely many unexposed subjects with propensity scores in the neighborhood of the propensity score of each exposed subject. Thus, $p_{0i}$ converges to $p_{1i}$ for all $i$. Since $H^a(t\mid X_c)$ is Lipschitz with respect to $X_c$ by Assumption \ref{finiteandlipschitzA_assump}, and it is easy to show that the propensity score $p$ is Lipschitz in $X_c$. It follows that $H^a(t\mid p)$ and therefore $S^a(t\mid p)$ are both Lipschitz with respect to $p$. Therefore, $\sum_{i=1}^m H^0(t \mid p_{0i})S^0(t \mid p_{0i})/\sum_{i=1}^m S^0(t \mid p_{0i})$ converges to $\sum_{i=1}^m H^0(t \mid p_{1i})S^0(t \mid p_{1i})/\sum_{i=1}^m S^0(t \mid p_{1i})$, which is the empirical marginal cumulative hazard based on the matched exposed subjects. Therefore, the discrepancy in the propensity scores due to matching vanishes as $m\to\infty$. 

We next prove the consistency of $\hat{H}^0_1(t)$. Recall that
\begin{align}
\hat{H}^0_1(t) &= \int_0^t\frac{\sum_{i=1}^m \rho_{1i}\hat{\phi}(\hat{p}_i)^{-1}dN_{0i}(u)}{\sum_{i=1}^m Y_{0i}(u)\rho_{1i}\rho_{0i}\hat{\phi}(\hat{p}_i)^{-1}}. \nonumber
\end{align}

If we had the full cohort instead of only the case-cohort samples, $H^0_1(t)$ can be estimated by the usual Nelson-Aalen estimator $\bar{H}^0_1(t)=\int_0^t \left[ \sum_{i=1}^{N_1} dN_{0i}(u) / \sum_{i=1}^{N_1} Y_{0i}(u)\right]$. Note that the summations in the estimator are over the full cohort exposed sample size $N_1$ due to matching. Let $\E_{p_1}(\cdot)$ denote the expectation with respect to the distribution of the propensity score in the exposed population. Then by Law of Large Numbers, $\bar{H}^0_1(t) \xrightarrow{p} \int_0^t \left[\E_{p_1}(dN_0(u)) / \E_{p_1} (Y_0(u)) \right]$.

First, consider the numerator of the integrand of $\hat{H}^0_1(t)$. Given that $\hat{\E}_{M_0}[\delta_{0i} \mid \hat{p}(X_{0i})]$ converges to $\E_{M_0}[\delta_{0i} \mid p(X_{0i})]$, $\hat{\phi}(\hat{p}_i)$ converges to $\phi(p_i)=\E_{M_0}[\delta_{0i} \mid p(X_{0i})] + [1-\E_{M_0}(\delta_{0i} \mid p(X_{0i})]\alpha_{b(i)}^{-1}$. Thus, $\hat{\phi}(\hat{p}_i)^{-1}$ converges to $\phi(p_i)^{-1}$ by Slutsky's theorem. On the other hand, conditioning on $M_0$ (i.e., $\xi_{0i}=1$ or $\delta_{0i}=1$) and given propensity score $p_i$, $\E_{M_0}(\rho_{0i} \mid p_i)=\E_{M_0}[\delta_{0i} \mid p(X_{0i})] + [1-\E_{M_0}(\delta_{0i} \mid p(X_{0i})]\E_{M_0}(\xi_{0i})\alpha_{b(i)}^{-1}=\pr[\delta_{0i}=1 \mid p(X_{0i}), M_0] + \pr[\delta_{0i}=0 \mid p(X_{0i}), M_0]\E_{M_0}(\xi_{0i})\alpha_{b(i)}^{-1}=\phi(p_i)$ since $\xi_{0i}=1$ whenever $\delta_{0i}=0$ in the case-cohort.

To facilitate the derivation, assume that $k$ unexposed subjects in the case-cohort sample are matched to each exposed subject. The one-to-one matching considered in this paper is a special case where $k=1$. Let $k^*_i$ be the number of unexposed subjects in the full cohort that would have been matched to exposed subject $i$ had there been no case-cohort design. Then $k^*_i=\sum_{j=1}^{k}\rho_{0ij}$ where $\rho_{0ij}$ is the case-cohort weight for the $j$-th unexposed subject in the $i$-th matched set. It follows that $\E(k^*_i\mid p_i)=k\phi(p_i)$. Under this setting, the numerator of the integrand of  $\hat{H}^0_1(t)$ becomes $\sum_{i=1}^m \rho_{1i}[k\hat{\phi}(\hat{p}_i)]^{-1}\sum_{j=1}^k dN_{0ij}(u)$. Since $\rho_{1i}=0$ for the exposed subjects who are not in the case-cohort and $\rho_{0i}=1$ whenever $dN_{0ij}=1$, the normalized numerator can be written as
\begin{align*}
& N_1^{-1} \sum_{i=1}^m \rho_{1i}\Big[k\hat{\phi}(\hat{p}_i)\Big]^{-1}\sum_{j=1}^k dN_{0ij}(u) \nonumber \\
= & N_1^{-1} \sum_{i=1}^{N_1} \rho_{1i}\Big[k\hat{\phi}(\hat{p}_i)\Big]^{-1}\sum_{j=1}^k \rho_{0ij}dN_{0ij}(u) \nonumber \\
= & N_1^{-1} \sum_{i=1}^{N_1} \rho_{1i}\Big[k\phi(p_i)\Big]^{-1}\sum_{j=1}^k \rho_{0ij}dN_{0ij}(u) + o_p(1). \nonumber 
\end{align*}
Given that $\rho_1 \indep [\rho_0, dN_0(u)] \mid p$, the expectation of the normalized numerator with respect to the distribution fo the propensity score in the exposed population:
\begin{align*}
& \E_{p_1}\left\{N_1^{-1} \sum_{i=1}^{N_1} \rho_{1i}\Big[k\phi(p_i)\Big]^{-1}\sum_{j=1}^k \rho_{0ij}dN_{0ij}(u) + o_p(1)\right\} \nonumber \\
= & \E_{p_1}\left\{N_1^{-1} \sum_{i=1}^{N_1} \E(\rho_{1i}\mid p_i)\Big[k\phi(p_i)\Big]^{-1}\E\left[\sum_{j=1}^k \rho_{0ij}dN_{0ij}(u)\mid p_i\right]\right\} + \E_{p_1}[o_p(1)] \nonumber \\
= & \E_{p_1}\left\{N_1^{-1} \sum_{i=1}^{N_1} \E(\rho_{1i}\mid p_i)\Big[k\phi(p_i)\Big]^{-1}\E\left[\sum_{j=1}^{k^*_i}dN_{0ij}(u)\mid p_i\right]\right\} + o(1) \nonumber \\
= & \E_{p_1}\left\{N_1^{-1} \sum_{i=1}^{N_1} \E(\rho_{1i}\mid p_i)\Big[k\phi(p_i)\Big]^{-1}\E(k^*_i\mid p_i)\E\left[dN_{0i}(u)\mid p_i\right]\right\} + o(1) \nonumber \\
= & \E_{p_1}\left\{N_1^{-1} \sum_{i=1}^{N_1} \E(\rho_{1i}\mid p_i)\E\left[dN_{0i}(u)\mid p_i\right]\right\} + o(1) \nonumber \\
= & \E_{p_1}(\rho_1)\E_{p_1}\left[dN_0(u)\right] + o(1). \nonumber
\end{align*}

Similarly, the normalized denominator of the integrand of $\hat{H}^0_1(t)$ under one-to-$k$ matching can be written as
\begin{align*}
& N_1^{-1} \sum_{i=1}^{N_1} \rho_{1i}\Big[k\hat{\phi}(\hat{p}_i)\Big]^{-1}\sum_{j=1}^k \rho_{0ij}Y_{0ij}(u) \nonumber \\
= & N_1^{-1} \sum_{i=1}^{N_1} \rho_{1i}\Big[k\phi(p_i)\Big]^{-1}\sum_{j=1}^k \rho_{0ij}Y_{0ij}(u) + o_p(1). \nonumber 
\end{align*}

Following the same derivation as for the numerator, we have
\begin{align*}
\E_{p_1}\left\{N_1^{-1} \sum_{i=1}^{N_1} \rho_{1i}\Big[k\phi(p_i)\Big]^{-1}\sum_{j=1}^k \rho_{0ij}Y_{0ij}(u) + o_p(1)\right\} = \E_{p_1}(\rho_1)\E_{p_1}\left[Y_0(u)\right] + o(1). \nonumber
\end{align*}

Taking the results of the numerator and denominator of the integrand of $\hat{H}^0_1(t)$ together and taking $k=1$, we have
\begin{align*}
 & \frac{\sum_{i=1}^m \rho_{1i}\hat{\phi}(\hat{p}_i)^{-1}dN_{0i}(u)}{\sum_{i=1}^m Y_{0i}(u)\rho_{1i}\rho_{0i}\hat{\phi}(\hat{p}_i)^{-1}} \nonumber \\
 = & \frac{N_1^{-1}\sum_{i=1}^{N_1} \rho_{1i}\hat{\phi}(\hat{p}_i)^{-1}dN_{0i}(u)}{N_1^{-1}\sum_{i=1}^{N_1} Y_{0i}(u)\rho_{1i}\rho_{0i}\hat{\phi}(\hat{p}_i)^{-1}} \nonumber \\
 =& \frac{\E_{p_1}\left[\rho_{1i}\hat{\phi}(\hat{p}_i)^{-1}dN_{0i}(u)\right] + o_p(1)}{\E_{p_1}\left[Y_{0i}(u)\rho_{1i}\rho_{0i}\hat{\phi}(\hat{p}_i)^{-1}\right]+o_p(1)} \nonumber \\
 =& \frac{\E_{p_1}(\rho_1)\E_{p_1}\left[dN_0(u)\right] + o(1) + o_p(1)}{\E_{p_1}(\rho_1)\E_{p_1}\left[Y_0(u)\right] + o(1)+o_p(1)}, \nonumber
\end{align*}
which is asymptotically equivalent to $\E_{p_1}[dN_0(u)]/\E_{p_1}[Y_0(u)]$. Thus, $\hat{H}^0_1(t)$ is asymptotically equivalent to $\int_0^t \left[\E_{p_1}(dN_0(u)) / \E_{p_1} (Y_0(u)) \right]$, which is asymptotically equivalent to the Nelson-Aalen estimator under the full cohort. Therefore, $\hat{H}^0_1(t) \xrightarrow{p} H^0_1(t)$.

Next, we establish the asymptotic normality of $\sqrt{m}[\hat{H}^0_1(t) - H^0_1(t)]$. 
\begin{align*}
    & \sqrt{m}[\hat{H}^0_1(t) - H^0_1(t)] \\
    & = \sqrt{m}\left[\int_0^t\frac{\sum_{i=1}^m \rho_{1i}\hat{\phi}(\hat{p}_i)^{-1}dN_{0i}(u)}{\sum_{i=1}^m Y_{0i}(u)\rho_{1i}\rho_{0i}\hat{\phi}(\hat{p}_i)^{-1}}- H^0_1(t) \right]\\
    & = \int_0^t \frac{m}{\sum_{i=1}^m Y_{0i}(u)\rho_{1i}\rho_{0i}\hat{\phi}(\hat{p}_i)^{-1}} \frac{1}{\sqrt{m}} \sum_{i=1}^m \rho_{1i}\hat{\phi}(\hat{p}_i)^{-1}dM_{0i}(u) + \sqrt{m} \int_0^t \frac{\sum_{i=1}^m \rho_{1i}\hat{\phi}(\hat{p}_i)^{-1}Y_{0i}(u) - \sum_{i=1}^m \rho_{1i}\rho_{0i}\hat{\phi}(\hat{p}_i)^{-1}Y_{0i}(u) }{\sum_{i=1}^m Y_{0i}(u)\rho_{1i}\rho_{0i}\hat{\phi}(\hat{p}_i)^{-1}} dH^0_1(u),
\end{align*}
where $dM_{0i}(u) = dN_{0i}(u) - Y_{0i}(u)dH^0_1(u)$ is a martingale by Doob Meyer decomposition.

Under Bernoulli sampling, $\xi$'s and therefore $\rho_1$'s and $\rho_0$'s are i.i.d. By the weak law of large numbers, $m^{-1}\sum_{i=1}^m Y_{0i}(u)\rho_{1i}\rho_{0i}\hat{\phi}(\hat{p}_i)^{-1} \xrightarrow{p} \pi_{01}(u)$ where $\pi_{01}(u)=\E[Y_{0i}(u)\rho_{1i}\rho_{0i}\hat{\phi}(\hat{p}_i)^{-1}]$. It follows that
\begin{align*}
    & \sqrt{m} [\hat{H}^0_1(t) - H^0_1(t)] \\
    & = \int_0^t \pi_{01}(u)^{-1}\frac{1}{\sqrt{m}}\sum_{i=1}^m \rho_{1i}\phi(p_i)^{-1}dM_{0i}(u) + \frac{1}{\sqrt{m}} \int_0^t \frac{\sum_{i=1}^m \rho_{1i}(1-\rho_{0i})\phi(p_i)^{-1}Y_{0i}(u)}{\pi_{01}(u)} dH^0_1(u) + o_p(1) \\
    & = J_1 + J_2 + o_p(1).
\end{align*}

Let us first consider the term $J_1$. Let $\mathcal{F}_0(u)$ be the filtration jointly generated by $M_{0i}(u)$, $\delta_{1i}$, and $\xi_{1i}$ ($i=1,...,m$). Then $M_{0i}(u)$ is adapted to $\mathcal{F}_0(u)$ and $\rho_{1i}$ is predicable with respect to $\mathcal{F}_0(u)$. Furthermore, $\pi_{01}(u)^{-1}$ and $\phi(p_i)^{-1}$ are bounded and deterministic, by the Martingale Central Limit Theorem, $J_1$ converges weakly to a mean-zero Gaussian process on $[0,\tau]$.

Now let us consider the term $J_2$. Under Bernoulli sampling, $\rho_{1i}(1-\rho_{0i})\phi(p_i)^{-1}Y_{0i}(u) dH^0_1(u)$ are i.i.d. random processes with bounded variation. Applying the Central Limit Theorem at each time $t$, we know that $J_2$ converges pointwise in distribution to a normal distribution with mean $\E(J_2)=\E[\int_0^t \rho_{1i}(1-\rho_{0i})\phi(p_i)^{-1}Y_{0i}(u)\pi_{01}(u)^{-1}dH^0_1(u)]=0$. By Example 2.5.4 in Vaart and Wellner \citep{wellner2013weak}, the class of indicator functions $\{Y_{0i}(u), u\in[0,t]\}$ is Donsker. Since $\rho_{1i}$, $1-\rho_{0i}$, $\pi_{01}(u)^{-1}$, $\phi(p_i)^{-1}$, and $dH^0_1(u)$ are all bounded, it follows that $\rho_{1i}(1-\rho_{0i})\phi(p_i)^{-1}Y_{0i}(u) dH^0_1(u)$ is also Donsker and therefore tight on $[0,t]$. Thus, $J_2$ converges weakly to a mean-zero Gaussian process.

Furthermore, since $\hat{H}^0_1(t) \xrightarrow{p} H^0_1(t)$,  we have $\E\{\sqrt{m}[\hat{H}^0_1(t) - H^0_1(t)]\}\xrightarrow{p}0$. By martingale property, $\E(J_1)=0$. It follows that $\E(J_2)\xrightarrow{p}0$. Thus,
\begin{align*}
    \mathrm{cov}(J_1,J_2) & \xrightarrow{p} \mathrm{E}(J_1 J_2) \\
    & = \mathrm{E}\big[J_1 \mathrm{E}(J_2\mid \mathcal{F}_0(t)) \big] \\
    & = \mathrm{E}\Bigg[J_1 \int_0^t \frac{1}{\sqrt{m}} \sum_{j=1}^m Y_{1i}(u)\rho_{1i}(1-\delta_{0i})\left(1-\frac{\E(\xi_{0i} \mid \mathcal{F}_0(t))}{\alpha_{b(i)}}\right)\phi(p_i)^{-1}\pi_{01}^{-1}(u) dA^0_1(u) \Bigg] \\
    & = 0.
\end{align*}
Therefore, $J_1$ and $J_2$ are asymptotically independent of each other. Putting the above results together, $\sqrt{m} [\hat{H}^0_1(t) - H^0_1(t)]$ converges weakly to a mean-zero Gaussian process on $[0, \tau]$. \qed

\section{Proof of Theorem 1}

By Theorem II.8.1 of Andersen et al. \citep{andersen2012statistical}, $\sqrt{m}[\hat{S}^a_1(t)-S^a_1(t)]$ is asymptotically equivalent to $-S^a_1(t)\sqrt{m}[\hat{H}^a_1(t)-H^a_1(t)]$ for $a=0,1$. By Lemma \ref{lemma1}, it follows that $\sqrt{m}[\hat{S}^a_1(t)-S^a_1(t)]$ converges weakly to a zero-mean Gaussian process uniform in $t\in[0,\tau]$ as $m\to\infty$. Thus, $\sqrt{m}[\hat{\mu}^a_1(\tau)-\mu^a_1(\tau)]=\int_0^{\tau}\sqrt{m}[\hat{S}^a_1(u)-S^a_1(u)]du$ converges in distribution to a zero-mean Gaussian random variable. Furthermore, since $\hat{S}^0_1(t)$ and $\hat{S}^1_1(t)$ are independent, it follows that $\sqrt{m}(\hat{\Delta}_{ATT}-\Delta_{ATT})=\int_0^{\tau}\sqrt{m}[\hat{S}^1_1(u)-\hat{S}^0_1(u)]du - \int_0^{\tau}\sqrt{m}[S^1_1(u)-S^0_1(u)]du=\int_0^{\tau}\sqrt{m}[\hat{S}^1_1(u)-S^1_1(u)]du - \int_0^{\tau}\sqrt{m}[\hat{S}^0_1(u)-S^0_1(u)]du$, which converges to a zero-mean Gaussian variable as $m\to\infty$. \qed

\end{document}